\documentclass[useAMS,usegraphicx,usenatbib]{mn2e}
\usepackage[a4paper,centering, totalwidth=520pt, totalheight=700pt]{geometry}
\usepackage{graphicx}
\usepackage{aas_macros}     
\usepackage{amsmath}
\usepackage{amssymb}
\usepackage{fixltx2e}[1999/12/01]

\newcommand{\Mass}{ \ensuremath{ h^{-1} M_{\odot}} }
\newcommand{\Msun}{ \ensuremath{M_{\odot}} }
\newcommand{\Mpc}{ \ensuremath{h^{-1} {\rm Mpc}} }

\newcommand{\kpc}{ \ensuremath{ h^{-1} {\rm kpc}} }

\newcommand{\treat}{{\tt T4PM}}
\newcommand{\tree}{{\tt Tree-PM}}
\newcommand{\pmesh}{{\tt PM}}
\newcommand{\proto}{{\tt Proto-Haloes}}
\newcommand{\halo}{{\tt Haloes}}
\newcommand{\nohalo}{ {\tt Not-Haloes}}

\usepackage[usenames,dvipsnames]{color}

\title[The Warm DM halo mass function below the cut-off scale]
{The Warm DM halo mass function below the cut-off scale.}

\begin{document}
\setlength{\topmargin}{-1.cm}

\author[Angulo et al.]{
\parbox[h]{\textwidth}
{Raul E. Angulo$^{1} \thanks{reangulo@stanford.edu}$, 
Oliver Hahn$^{2}\thanks{hahn@phys.ethz.ch}$, 
Tom Abel$^{1}\thanks{tabel@stanford.edu}$.} \vspace*{6pt} \\
\\
$^1$  Kavli Institute for Particle Astrophysics and Cosmology,\\
Stanford University, SLAC National Accelerator Laboratory, Menlo Park, CA 94025, USA \\
$^2$ Department of Physics, ETH Zurich, CH-8093 Z\"urich, Switzerland.}
\maketitle

\date{\today}
\pagerange{\pageref{firstpage}--\pageref{lastpage}} \pubyear{2013}
\label{firstpage}

\begin{abstract} 
Warm Dark Matter (WDM) cosmologies are a viable alternative to the 
Cold Dark Matter (CDM) scenario. Unfortunately, an accurate scrutiny
of the WDM predictions with N-body simulations has proven difficult 
due to numerical artefacts. Here, we report on cosmological simulations that,
for the first time, are devoid of those problems, and thus, are able to 
accurately resolve the WDM halo mass function well below the cut-off. 
We discover a complex picture, with perturbations at different evolutionary
stages populating different ranges in the halo mass function. 
On the smallest mass scales we can resolve, identified objects are typically centres of  
filaments that are starting to collapse. On intermediate mass scales, objects 
typically correspond to fluctuations that have collapsed and are in the process 
of relaxation, whereas the high mass end is dominated by objects similar to 
haloes identified in CDM simulations.
We then explicitly show how the formation of low-mass haloes is 
suppressed, which translates into a strong cut-off in the halo mass function.
This disfavours some analytic formulations that predict a halo mass function
that would extend well below the free streaming mass.
We argue for a more detailed exploration of the formation of
the smallest structures expected to form in a given cosmology, which,
we foresee, will advance our overall understanding of structure formation.
\end{abstract}

\section{Introduction} 

Since the 1980s, Warm Dark Matter (WDM) has been an attractive alternative to
Cold Dark Matter (CDM) as the main gravitating component in the Universe. 
For a long time, however, WDM was disfavoured compared to CDM due to, in part, the  
additional free parameter required (the mass of the DM particle). This has 
changed recently, and WDM has again attracted the attention of the cosmological
community as a viable and competitive cosmological model. 
Traditionally, the gravitino was favoured as the hypothetical particle that 
could serve as WDM \citep[e.g.][and references therein]{Moroi1993},
while, more recently, the interest has focused on the sterile neutrino 
\citep[see e.g.][for a recent review]{Boyarsky2009a}.

The reason for the recent revitalisation of WDM is that a suitable $\Lambda$WDM 
model could reproduce all the successes of $\Lambda$CDM on large scales, and, 
in addition, it may alleviate the alleged tension between the CDM model and 
some observations. Such measurements concern mainly mass scales of 
$\sim 10^{10}\Msun$ and include: the dynamics of Milky-Way satellites 
\citep{Boylan-Kolchin2011,Lovell2012}, the velocity function
of HI selected galaxies \citep{Zavala2009}, and the abundance of low
mass galaxies at low and high redshifts \citep{Menci2012}. 

The key feature of WDM cosmologies that distinguishes them from CDM, is the lack 
of initial small-scale density 
fluctuations. Before recombination, WDM particles have relatively high velocities
(set when these particles become non-relativistic), which allows them to travel
further than a ``free-streaming'' distance $R_{\rm fs}$. Thus, particles move out of 
overdense regions of size $R_{\rm fs}$ and smaller, and therefore 
inflation-generated small-scale density and potential perturbations are washed 
out. This dissipation is captured as a strong suppression of the mass transfer 
function below a ``cut-off'' scale \citep[cf.][]{Bond1983}. This difference relative 
to CDM is expected to affect the abundance of collapsed objects (DM haloes) below 
the cut-off mass scale\footnote{Note that the free-streaming
scale {\it today} is much smaller than that at early times. Thus, scales below the
cut-off scale in the transfer function are, in principle, able to collapse
gravitationally, but not those below today's value of the free-streaming scale. } 
\citep[e.g.][]{Schneider2012,Menci2012,Benson:2013}, and also the
internal properties of haloes \citep[e.g.]{Colin2000}. In particular, it is expected the abundance of low-mass haloes to
be suppressed, and the halo density profiles to be less concentrated. In addition,
the formation time, as well as the actual halo formation mechanism is modified,
as a larger fraction of haloes form directly from the collapse of filaments.
Additionally, phase correlations of the overall density field are expected to
be related to the mass of the DM particle \citep{Obreschkow2013}.

Besides the obvious cosmological interest in WDM scenarios, this
class of models also offers an interesting test case from a 
theoretical point of view, especially for cosmological simulations. Free-streaming provides
a small-scale limit to the structure formation problem, thus, in 
principle, it is possible to capture the full hierarchy of objects
expected to be relevant during the formation of a halo. This can
provide clues as to how hierarchical structure formation proceeds
in general, and its connection to cosmological parameters and the spectrum
of density fluctuations. In contrast, this is less direct in CDM 
since there are always structures and interactions not resolved by any 
given mass and force resolution at late times.

Unfortunately, an accurate numerical scrutiny of the predictions of a 
WDM model has proven difficult. Despite a multitude of attempts, N-body 
simulations, and analytical frameworks, have not been able to decisively 
quantify even one the simplest properties of the nonlinear field: 
the abundance of DM haloes. Cosmological WDM simulations show
the dominant presence of an artificial population of low-mass DM haloes. This
phenomenon originates in the numerical (unphysical) fragmentation of filaments, 
and it exceeds by far the population of real WDM haloes on scales near the cut-off scale
~\citep[see e.g.][]{Avila-Reese:2001,Bode2001,Wang2007,Melott2007}.
Although higher mass resolution helps to somewhat reduce this problem \citep[e.g.][]{Schneider2013}, the mass
scale dominated by artificial fragmentation changes only weakly with an increase 
of the  mass resolution. 
Besides the obvious impact of artificial fragmentation in the halo mass function, 
it is possible that they also affect the internal properties of
DM haloes, which grow by accreting these objects.
Finally, it is currently unknown what type of structures, if any, inhabit below 
the cut-off mass scale. The latter is relevant when modelling galaxy formation 
in WDM scenarios. 

Similarly, recent analytic approaches, based on modifications of the Extended
Press-Schechter theory \citep{Press1974,Lacey1993}, produce predictions that 
differ by orders of magnitude from each other in the way in which the halo mass 
function is suppressed in WDM 
~\citep[compare e.g.][]{Smith:2011,Benson:2013}. The
reason for this is mainly uncertainties in the formulation of traditional
excursion sets, together with possible modifications to the shape of the barrier 
for collapse near the cut-off scale.

In this paper, we revisit the issue of the halo mass function in WDM cosmologies
and propose an answer to the questions posed above. In order to overcome previous
limitations, we carry out a suite of high-resolution cosmological N-body simulations 
that feature a recently developed method to compute gravitational forces \citep{Hahn2012}.
This new technique allows us to alleviate some of the long-known problems
originating from employing excessive force resolution compared to the
mass resolution in the simulations \citep{Efstathiou1981,Centrella1988,Melott1989,Splinter1998}, and also to reduce
discreteness noise in the large-scale density and tidal fields, which
we find are key players in causing artificial fragmentation. 

With these
tools at hand, we are able to robustly compute for the first time the WDM 
halo mass function at, and below, the cut-off mass scale. Our simulations 
unveil systems of different characteristics, and at different stages of 
formation, populating different mass ranges in the halo mass function. 
Explicitly, well below the cut-off scale we find dense filaments and sheets
that have started to collapse into 3D systems. At larger mass scales we 
find proto-haloes, systems that are collapsed but still in the process of virialisation.
Only above the cut-off scale we observe systems traditionally regarded as DM
haloes. When we consider only halos, We observe a strong suppression of the 
halo mass function, together with a cut-off on small masses, however. However, 
due to the wealth of different types of
structures, the position of the cut-off depends on the actual halo definition 
one wishes to adopt.

The structure of the paper is as follows: In \S2 we provide details of the
simulation techniques and the construction of the halo
catalogues. In \S3 we present our results, focusing on the abundance of collapsed
objects, the WDM halo mass function, and exploring the characteristics of objects 
located at different mass ranges in the halo mass function. 
We then discuss our findings and possible implications. Finally, in \S4 we provide a
summary of our work along with directions for future work.

\section{Numerical Tools}

In this section, we describe the numerical simulations we use to study dark matter
haloes in a WDM scenario. We also describe our halo identification procedure and
the construction of the group catalogues.

\subsection{Initial conditions}

We start by computing the power spectrum of density fluctuations using the
fitting formulae of \cite{EisensteinHu1999}. We adopt a set of cosmological parameters
consistent with the published measurements of the WMAP7 data release
\citep{Komatsu2010}. Explicitly: $\Omega_m = 0.276$, $\Omega_{\Lambda} =
0.724$, $\Omega_{b} = 0.045$, $h = 0.703$, $\sigma_8 = 0.811$ and spectral
index $n_s = 0.96$. Note we set the normalisation of the power spectrum 
according to $\sigma_8$ via a CDM spectrum, so that the amplitude of 
fluctuations on large-scales is independent of the WDM particle mass.

We then incorporate the effects of a thermally produced warm dark
matter particle on the transfer function following the fitting formula of
  \cite{Bode2001}.\footnote{See also \cite{Viel2005} who adopt a very
  similar fitting function but indicate a simple way in which
  (non-thermal) sterile neutrinos can be accounted for as well by a
  change to an effective thermal mass
  \citep[cf. also][]{Colombi1996}.} Explicitly, a fit to the WDM transfer 
  function of density perturbations is given by:

\begin{equation}
 T_{\rm WDM}(k) = T_{\rm CDM}(k) \left[1 + (\alpha\,k)^2\right]^{-5.0},
\end{equation}

\noindent with

\begin{eqnarray} 
\alpha & = & 0.05 \left(\frac{\Omega_m}{0.4}\right)^{0.15} \quad\times\\ 
& \times & \left(\frac{h}{0.65}\right)^{1.3}  
\left( \frac{m_{\rm dm}}{1\,{\rm keV}}  \right)^{-1.15}  \left(\frac{1.5}{g_X}\right)^{0.29}\,h^{-1}{\rm Mpc}, \nonumber
\end{eqnarray}

\noindent where $m_{\rm dm}$ is the DM particle mass (or the effective
sterile neutrino mass), in units of ${\rm keV}$, 
and $g_X$ is the number of degrees of freedom that the particle contributes to the number density which
is $3/2$ in our case. 

In this paper we will explore the case where $m_{\rm dm} = 250$ eV. 
Our choice is inconsistent with current constraints placed by observations of the Ly-$\alpha$ 
forest power spectrum (which set a lower limit at the order of keV \citep{Viel2005,Boyarsky2009}). However, a low $m_{\rm dm}$ has the advantage
of allowing us to resolve structures at redshift zero much below the cutoff 
mass scale using only relatively modest mass resolution and computational resources.
In addition, our results can be readily extended and generalised to others DM
particle masses.

For our choice of cosmological parameters and WDM particle mass: $\alpha = 0.26\,h^{-1}{\rm Mpc}$, which translates into a free-streaming mass-scale 
\begin{equation}
M_{\rm fs} =\frac{4{\rm \pi}}{3}\bar{\rho}\left(\alpha/2\right)^3\simeq7\times10^{8}\,h^{-1}{\rm M}_\odot
\end{equation}
and a half-mode mass-scale \citep[cf.][]{Colin2008,Schneider2012}
\begin{equation}
M_{\rm hm}\simeq 4.3\times10^{3}\,M_{\rm fs}\simeq3.0\times10^{12}\,h^{-1}{\rm M}_\odot.
\end{equation}
This mass-scale is where a deficit in the mass-function compared to CDM is expected
to be a factor of two. 

Using the WDM primordial power spectrum discussed above
and the {\tt MUSIC} code \citep{HahnAbel2011}, we then create the
initial particle configuration at $z = 63$ for our numerical
experiments. We do this by perturbing a particle distribution,
initially arranged in a regular cubic lattice, according to the
Zel'dovich approximation. 

We note that we have not attempted to explicitly include thermal velocities (on top of
gravitationally induced velocities) in our initial conditions, since it is 
both negligible for the results of this paper as well as unclear how
a proper implementation would proceed. The RMS velocity of the microscopic 
WDM particles following the Fermi-Dirac distribution (for thermally produced
warm dark matter) is given at redshift $z$ by \citep{Bode2001} 
\begin{eqnarray}
\sigma_v & \simeq & 0.043\left(1+z\right)\quad \times \\
& \times & \left(\frac{\Omega_m}{0.3}\right)^{1/3}\left(\frac{h}{0.65}\right)^{2/3}\left(\frac{1.5}{g_X}\right)^{1.3}
\left(\frac{1\,{\rm keV}}{m_{\rm dm}}\right)^{4/3}\,{\rm km}\,{\rm s}^{-1}.\nonumber 
\end{eqnarray}
For the $m_{\rm dm}=0.25\,{\rm keV}$ chosen in this paper, we find 
$\sigma_v = 0.28\,{\rm km}\,{\rm s}^{-1}$ at $z=0$. Clearly, a value considerably
smaller than those arising due to nonlinear structure formation. For instance, 
the virial velocity of a typical DM halo in our simulations is 100-1000 ${\rm km}\,{\rm s}^{-1}$. 
Of course, this RMS velocity can be of relevance in the most central parts of
the haloes, determining details of the phase-space density there
\citep[e.g.][]{Dalcanton2001}, as well as the thickness of caustics \citep[e.g.][]{White2009}.

Although the RMS velocity corresponds to a {\em microscopic} value, it is sometimes 
regarded as a macroscopic one, and implemented in N-body simulations as random 
kicks of simulation particles \citep[e.g.][]{Colin2008,Maccio2013}.
We emphasise, however, that this is simply an Ansatz and that simulation particles 
represent a coarse-grained phase-space distribution, thus each of them averages over a 
statistical ensemble with a negligibly small dispersion around this mean. 
On the contrary, a kick to a single simulation particle is equivalent to a locally coherent 
motion of a large ensemble of actual WDM particles. This leads to a velocity 
spectrum inconsistent with the results 
of linear perturbation theory \citep[][]{Colin2008}, and it is  clear that
the evolution of this numerical setup is not equivalent to the evolution of
the fine-grained distribution function. Therefore, neglecting this dispersion 
is a very good, as well as convenient, approximation when one is only concerned 
with the mass of DM haloes, as it is in our case.

\subsection{Gravitational Evolution}
\label{sec:evolution}
We perform a series of cosmological N-body simulations evolving $1024^3$
particles inside a cubical region of $L=80\,\Mpc$ a side. For our choice of
cosmological parameters, each of these simulation particles has a mass equal to
$3.65\times10^7\,\Mass$. This mass resolution and volume is sufficient to have
a fair sample of haloes located at the half-mode mass, which is resolved 
with almost $100,000$ particles.

We evolve simulation particles using a memory-efficient version of the
P-Gadget3 code \citep{Springel2005b}, which was originally developed
and optimised for the Millennium-XXL project \citep{Angulo2012a}. In
this code we have implemented three different methods to compute
gravitational forces. In the remainder of the paper we refer to them
as {\tree}, {\pmesh} and {\treat}, and are described in the following:

\begin{itemize}
\item[1.] {\bf \tree}: This method corresponds to the standard numerical
configuration followed in state-of-the-art calculations. Long-range
interactions are calculated using a PM method \citep{Hockney1981}, whereas
short-range forces are calculated using a multipole expansion of the force
field together with a Tree algorithm \citep{Barnes1986}. In order to reduce
two-body scattering and binary particle systems (among other artefacts), forces
need to be softened on small scales. We do this in our runs using a
Plummer-equivalent softening length equal to $5\,\kpc$.

\item[2.] {\bf \pmesh}: Here, gravitational forces are only given by the PM
method, i.e. we compute the gravitational potential field on a grid by solving
the Poisson equation using Fourier methods. In our runs we use a grid of
$2048^3$ points and forces are Gaussian smoothed on scales roughly equal to
twice the grid size, $2\Delta x = 80\,\kpc$.  This length scale matches the
mean interparticle separation. As argued by \cite{Angulo2013}, this numerical
configuration suppresses undesired collisionality of the N-body system, and
it is particularly successful (compared to a {\tree} run) in following accurately
the gravitational interaction of baryons and DM.

\item[3.] {\bf \treat}: This method is an implementation of the algorithm
proposed by \cite{Hahn2012}. In short, a Delaunay triangulation of the Lagrangian
particle distribution defines a phase-space element (a tetrahedron) that can be
reconstructed at any desired later time to reconstruct the respective 
density field. At all times, the density of each tetrahedron is
assumed to be uniform. In practice, we represent the contribution of each 
tetrahedron to the total mass field using $4$ 
virtual particles (they carry mass, but do not interact directly with the fluid) whose 
spatial distribution matches the monopole and the quadrupole of the parent tetrahedra. 
We deposit these particles onto a $4096^3$ mesh with CIC interpolation and compute 
forces using the PM method, as described above. The spatial resolution of these runs
is $40\,\kpc$, twice as high as in the {\pmesh} case. 
\end{itemize}

\begin{figure} 
\includegraphics[width=8.5cm]{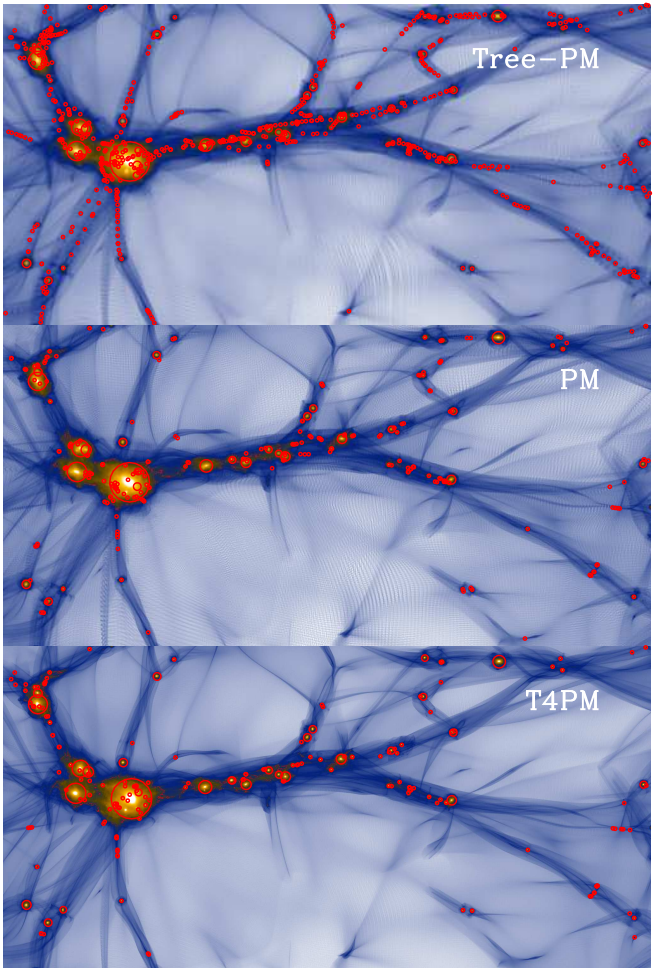} 
\caption{ Images of the cosmic density field at $z=0$ in a WDM cosmology
with $m_{\rm dm} = 0.25 keV$, as predicted by the three
different methods to compute gravitational forces. Each image displays a
projection of a $20\,\Mpc$ thick slab, of a region of size $80\times40\,\Mpc$,
which focuses on the most massive halo present in the simulation volume.
Additionally, in the three panels all haloes with mass $M_{200} > 1\times10^{10}\,\Mass$ are
displayed using red circles whose radii equal the haloes' virial radii, $R_{200}$.
Note
that the different methods show different amount of spurious fragmentation, 
which results in sequences of haloes aligned along filaments.
\label{fig:field}} 
\end{figure}

In Fig.~\ref{fig:field} we show the density projection of a subregion of our
WDM simulations, as computed by the three methods described above. We will
analyse this figure in detail in the next section, but it is readily apparent
that all three runs display the same large-scale structure, but they differ
on small scales. 

Additionally, we have carried 
out a set of CDM N-body simulations, with identical
force and mass resolution as the WDM runs. The initial conditions of all 
simulation boxes were generated using identical white-noise fields, which 
simplifies their comparison.
 
\subsubsection{Computational Performance} 
 
The computational resources required for the three methods are similar
but differ systematically. For our WDM runs, the {\tree}, {\pmesh} and
{\treat} runs require about $3000$, $1000$ and $3500$ CPU-hours,
respectively.  For the {\tree} run, $\sim50\%$ of the time was
employed in the construction and walking of the tree, while the
overhead of the {\treat} respect to the {\pmesh} method is caused by
our on-the-fly calculation of the initial tessellation together with
the position calculation and depositing of the mass-carriers particles. The peak
memory consumption of the {\treat} run is $\sim600$Gb, to be compared
with $\sim150$Gb employed by the {\pmesh} run. The difference is
dominated by an extra set of pointers needed to recover the initial
connectivity of the phase-space tessellation. We remark that the extra
factor of $\sim4$ is small considering that we effectively have
$24$ times more
particles\footnote{Each particle contributes to six distinct tetrahedra, and
each tetrahedra is represented by four particles.} representing the density and force field. For a dramatic 
increase in force accuracy it hence seems very worthwhile to afford 
this additional cost in memory and run time. In addition, our implementation 
is suboptimal in terms of memory consumption: it is possible to carry out the {\treat} run
with a memory footprint identical to that of {\pmesh}, at the cost of slightly 
more CPU time.
 
\subsection{Dark Matter Haloes}

For each of our simulations, we produce {\it on-the-fly} friends-of-friends
\citep[FoF,][]{Davis1985} halo catalogues. We use a non-standard linking length parameter of
$b=0.05$ times the mean inter-particle separation, keeping objects with $20$ or
more particles. This unusual choice of $b$ (as
compared to $b=0.2$) is required to avoid large FoF haloes percolating the
cosmic web. We will return to this point below.

For each FoF halo, we compute a spherical-overdensity (SO) mass, taking 
the centre of mass of the parent FoF group as the SO centre. We define the halo boundary
as the sphere of radius $R_{200}$, which contains a mean density of $200$ times
the critical density, $\rho_{crit}$. Therefore, the mass of the halo is $ 
M_{200} = \frac{4}{3}\,{\rm \pi}\,R_{200}^3\,200\,\rho_{crit}$.

We discard substructures from our catalogues
whose $R_{200}$ spheres overlap with that of a more massive halo.
At $z=0$ this procedure finds $8359$, $3422$ and $2916$ objects with mass
$M_{200}>10^{10}\Mass$ in the WDM {\tree}, {\pmesh}, and {\treat} runs,
respectively. These are a factor of $15-40$ smaller than in a CDM {\tree} run,
where we detect $127'133$ structures.  In Fig.~\ref{fig:field}, we overplot this
halo catalogue on top of the dark matter density field.

\begin{figure} 
\includegraphics[width=4.3cm]{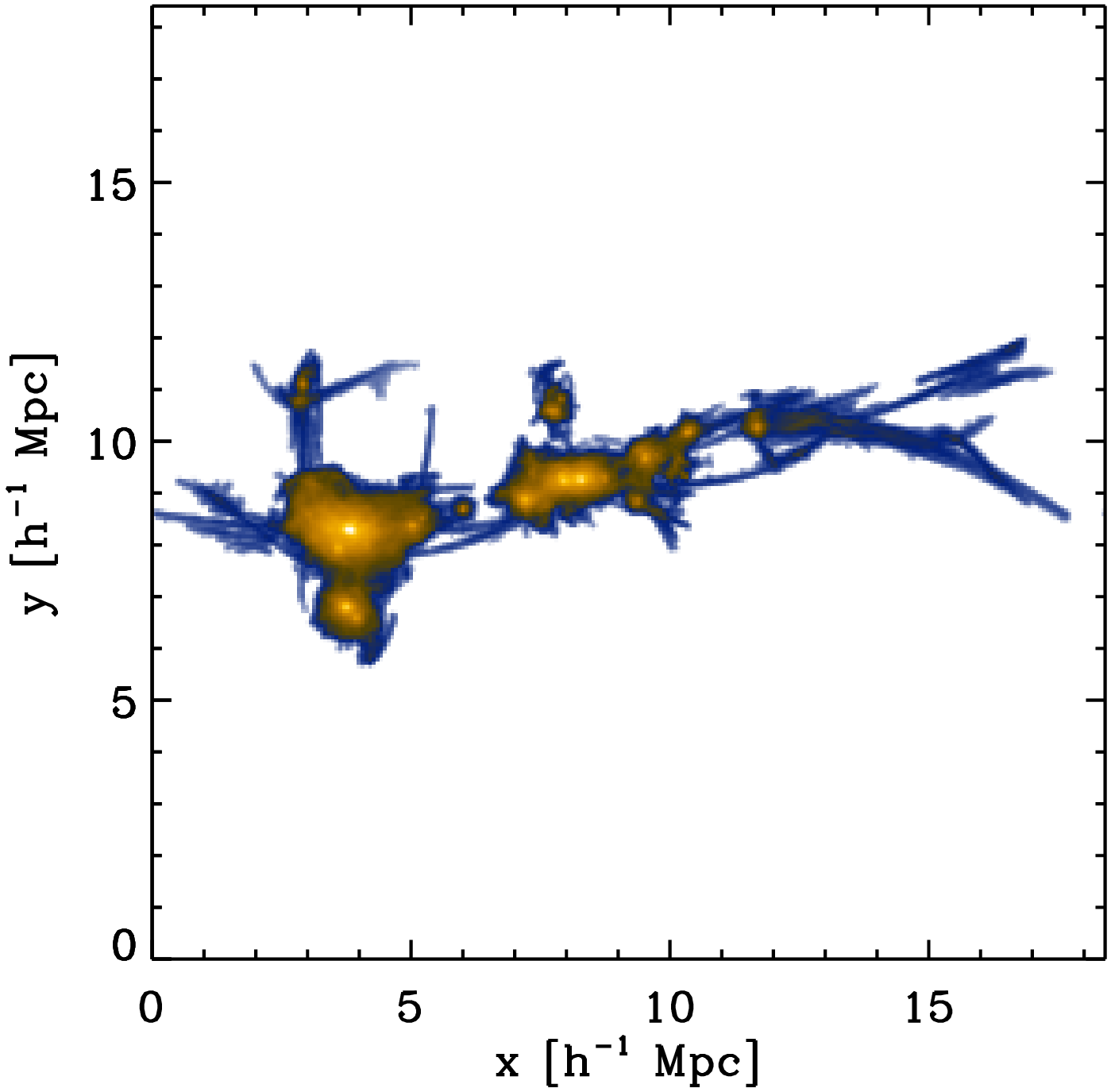} 
\includegraphics[width=4.3cm]{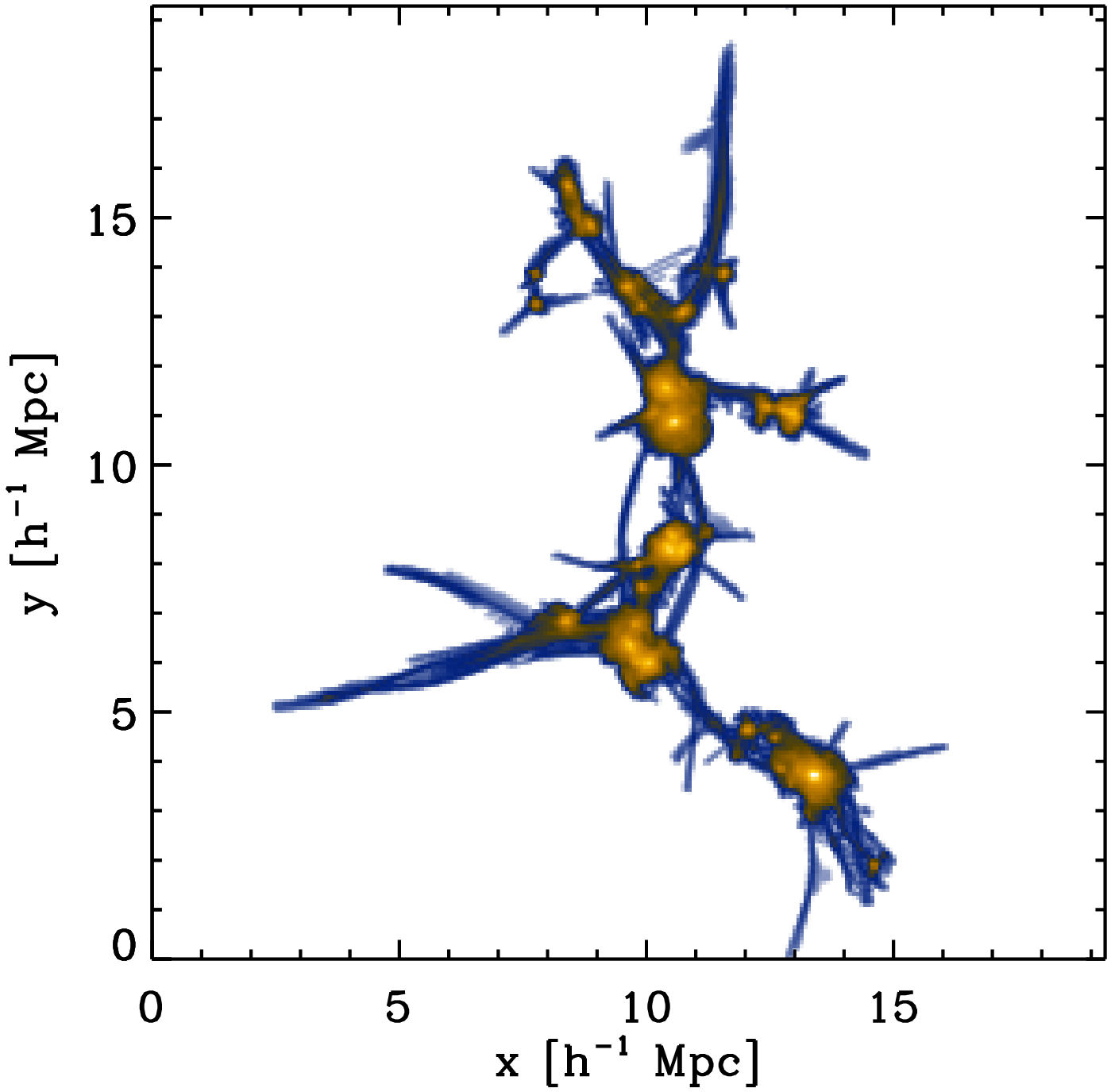} \\
\includegraphics[width=4.3cm]{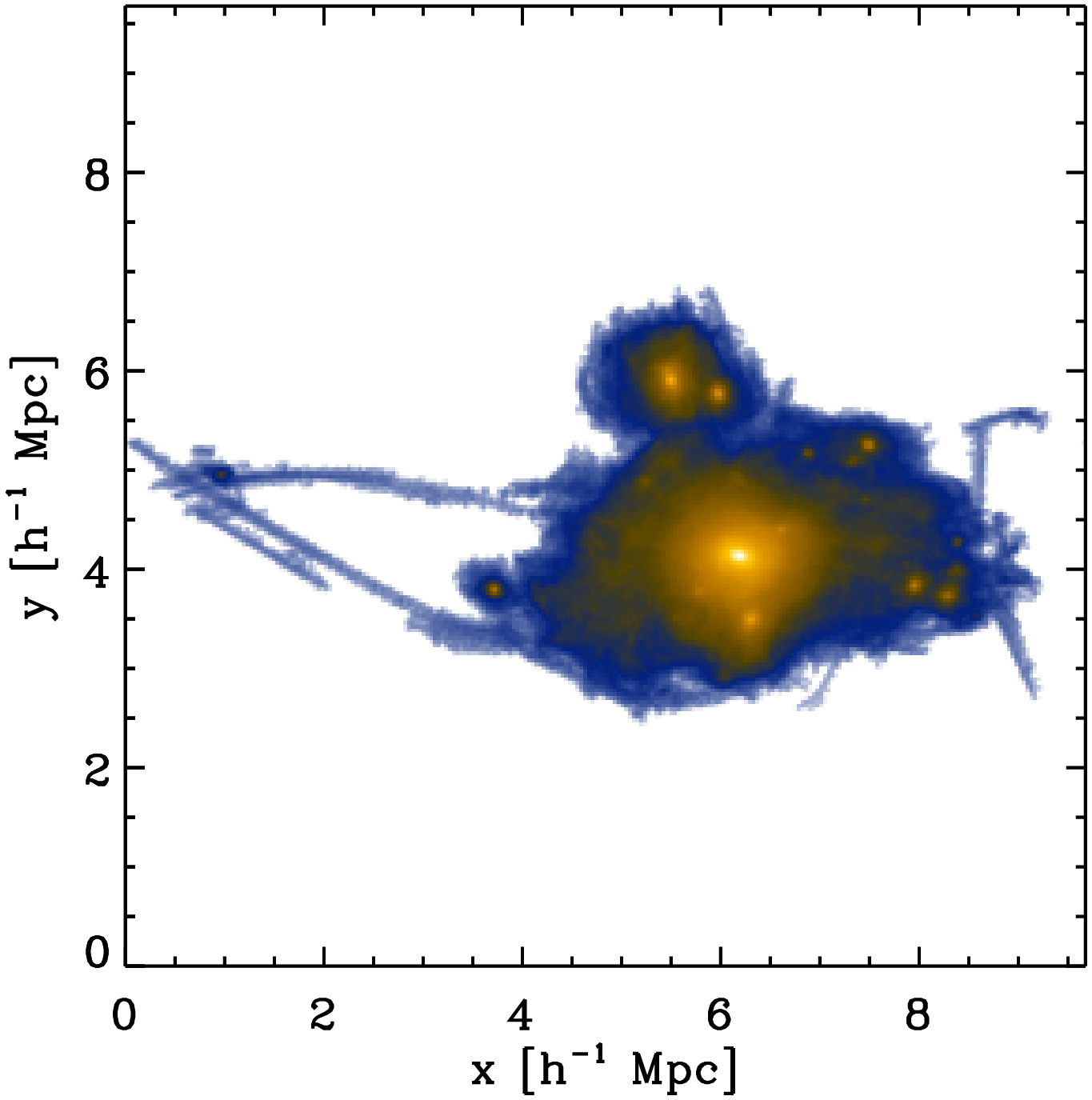} 
\includegraphics[width=4.3cm]{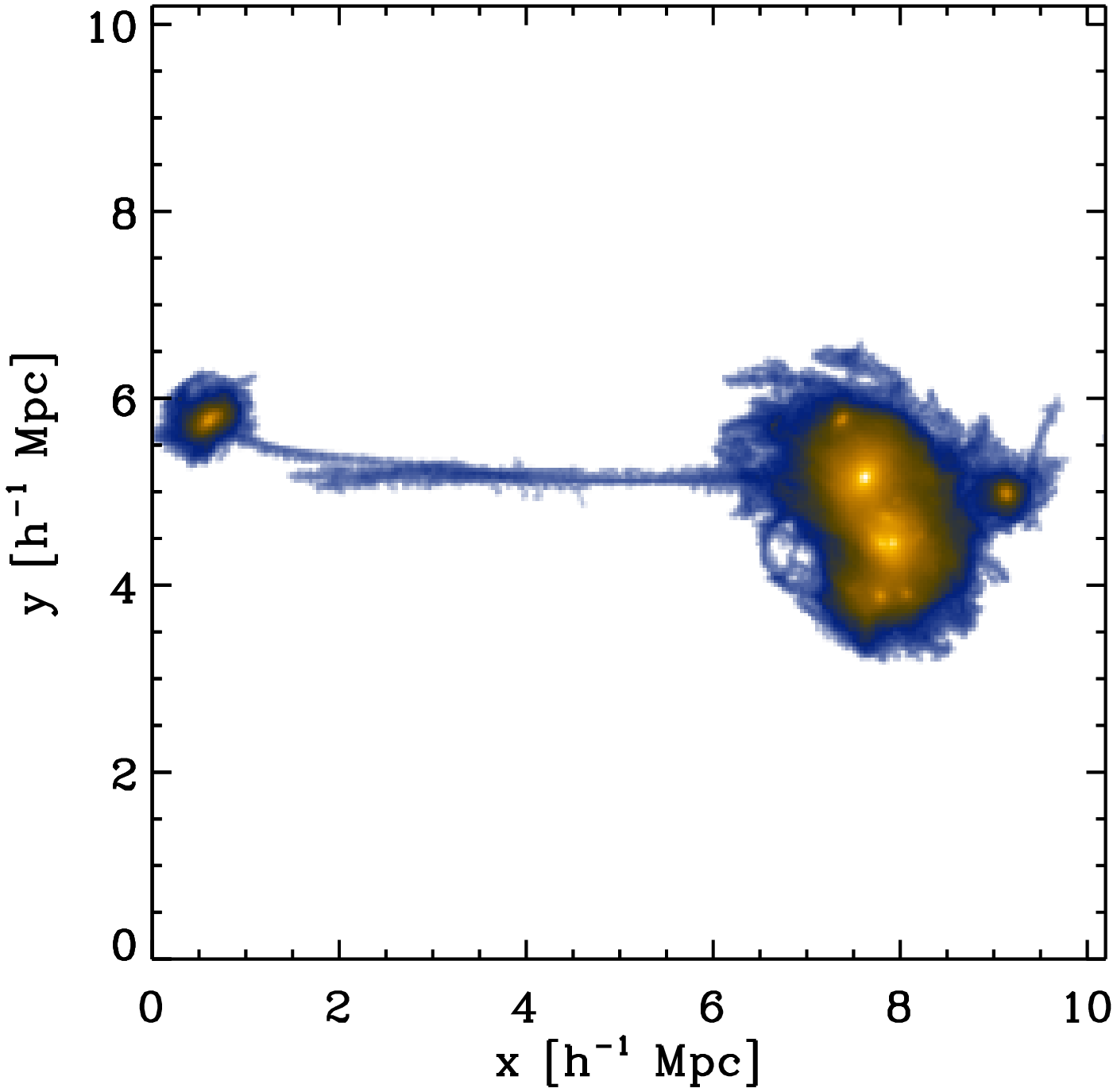} 
\caption{The 'standard' linking length $b=0.2$ selects large parts of a WDM
simulation once the forces are captured accurately enough that filaments do not
artificially fragment. Density projections of the particles belonging to the
two most massive FOF-``haloes'' in our WDM {\treat} simulation of a $250$eV dark
matter model. Objects at $z=1$ (top row) and $z=0$ (bottom row) are shown.
These haloes have a mass of $2.7\times10^{14}\Mass$ and $1.6\times10^{14}\Mass$
($z=1$), and of $6.4\times10^{14}\Mass$ and $2.8\times10^{14}\Mass$ ($z=0$).
\label{fig:fofcrash}}
\end{figure}

In passing, we note that even though the FoF algorithm with the standard choice, $b=0.2$,
works satisfactorily for the {\tree} run, it fails to deliver a 
reasonable halo catalogue for the other two cases. While in the {\tree} 
case the filaments are broken into small haloes, in the {\treat} and {\pmesh} runs 
a strong artificial fragmentation is absent (as
can be seen in Fig.~\ref{fig:field}), thus filaments and sheets
are more homogeneous with sharp dense cores: genuine two and one
dimensional dense structures exist. The FoF algorithm links these filaments to
dark matter haloes located at their ends, and with other nearby haloes. 
We show examples of this problem in Fig.~\ref{fig:fofcrash} where we display
a projection of the particles associated with the two most massive FoF haloes 
at $z=1$ (top row) and $z=0$
(bottom row), as identified in the {\treat} runs. The largest $b=0.2$ FoF
structure at $z=0$ has a mass of $6\times10^{14}\Mass$ and spans almost $10\,\Mpc$. At
$z=1$, the failure of the FoF algorithm is even worse: the biggest FoF halo
spans almost one quarter of the simulation box size! 

In order to avoid such problems, we employed a small linking length that ensures 
that only local high density peaks are selected as the starting point for our 
SO halo catalogues. We tested that the resulting SO halo mass function was 
insensitive to small changes in $b$ about our preferred value of $0.05$. However, 
for values approaching $b=0.2$, the mass function agrees only with the cases 
with smaller linking lengths at the highest mass end. On any other mass scale, 
it shows a notorious deficit of structures. This is because a large fraction of 
small haloes are artificially linked to form a single larger FoF structure, and 
thus are not present in our list of SO candidate haloes.

\section{Results}

The main goal of this paper is to quantify the abundance of haloes
expected in WDM cosmologies, especially below the cut-off scale. An accurate
account of this is important, firstly, to establish robustly the 
predictions of WDM which can then be tested against observational
data, and secondly, to understand more generally the collapse and
assembly of DM haloes in the presence of a resolved cut-off scale 
in the perturbation spectrum. This in turn can help to understand 
the formation and properties of micro-haloes expected for some
CDM particle candidates (e.g. the neutralino). In addition, these
cosmologies offer a test of the methods and implementations of
N-body simulations.

\subsection{Halo abundance -- dependence on the numerical method}

Previous numerical simulations have not been able to explore the
cut-off mass scale because it is dominated by a population of 
low-mass haloes aligned within filaments. This phenomenon has been 
reported in numerical simulation for decades: including early works 
\citep[e.g.][]{Melott1989,Avila-Reese:2001,Bode2001, Knebe2003}, and 
also recent state-of-the art runs \citep{Wang2007, Lovell2012,
Schneider2012}.

Initially, it was not clear whether a real and physical fragmentation of
filaments could be in place, or if it has its origin in numerical inaccuracies.
This has settled recently, and there is a consensus that these haloes
are numerical artefacts.  Evidence for this is that their spatial distribution
is closely related to the initial unperturbed particle load, and that their abundance
changes (albeit slowly) with mass resolution (in fact $\propto m_p^{1/3}$). 
In particular, \cite{Wang2007} have
analysed this problem in detail and concluded that their presence is caused by
non-zero small-scale fluctuations of the 1D-projected density
field. This is related to warnings of \cite{Melott1989} about using
excessive force resolution compared to the mass resolution as it leads
to fragmentation also in 2D cosmological simulations. This is indeed the regime in which
state-of-the-art simulations are carried out: the typical force resolution
is set to a value $10-100$ times smaller than the mean interparticle separation.

The numerical nature of the fragmentation is also illustrated in
Fig.~\ref{fig:field}, where we show a density projection of a $20\,\Mpc$ thick
slab through our three simulation boxes at $z=0$. The
visualization technique is identical for all three panels, and corresponds to a
CIC density (for the {\treat} run, we project the flow tracers, not the $24$
times more abundant mass carriers); thus any difference is a result of real
discrepancies in the spatial distribution of particles. In these images, we 
overplot the halo distribution over the underlying dark matter field. We 
display only haloes with SO mass $M_{200} > 2\times10^{10}\Mass$, which is 
the resolution limit of our simulations, as we will discuss below. It is
straightforward to see that all three runs, which use different methods to 
compute gravitational forces, display the same large-scale structure while, however, 
differences exist on small scales.

In the top panel, we display results obtained with the most commonly used method to compute
gravitational forces (c.f.\S2.1), labelled as {\tree}. Fragmentation of
filaments into small clumps is clearly visible in several places, for instance,
in the two filaments located in the lower half of the image. These clumps are
indeed very dense and are identified as haloes by our FoF-SO algorithm, and are
thus highlighted by red circles.  In the middle panel, which shows the {\pmesh}
simulation,  forces are effectively softened below the mean inter-particle
separation and low-mass haloes aligned with the filaments are considerably less
abundant. In the bottom image, displaying the {\treat} run, artificial fragmentation 
virtually does not exist! Even though this run has a force resolution 
twice as high as in the {\pmesh} case.  

Therefore, we see that the fragmentation of filaments is closely related 
to the force calculation, or more precisely, to the combination 
of force and mass resolution. We note that, in these simulations, the crucial difference 
is not the actual method to compute forces (e.g. a PM versus a Tree+PM), 
but the chosen force resolution for a given mass resolution. We have explicitly 
tested this assertion by varying the size of the mesh in the {\pmesh} run. 
An excessive
force resolution causes local minima in the global potential around simulation particles,
which eventually grow and accrete neighbouring particles. 
In this sense the {\treat} method has the advantage of smoothing
these minima since it provides a smoother representation of the mass field 
\cite[see e.g.][]{Kaehler2012} and thus more accurately captures the smooth
but dense structure of the density field in regions of strong anisotropic compression. 
This was already qualitatively seen by 
\cite{Hahn2012}, who found that the {\treat} method suppresses artificial 
fragmentation in WDM scenarios. The advantage of {\treat} has the price that 
the density in the
inner regions of haloes is overestimated. This is because the evolution of
highly distorted Lagrangian phase-space elements in regions of strong mixing
cannot be represented correctly by the piecewise linear approximation to the 
distribution function that is only tracked by the Lagrangian motion of the
particles.
In principle this limitation can be overcome by an adaptive mass refinement 
(Hahn, Angulo \& Abel, 2013, in prep.). Nevertheless, this limitation has 
 a very minor effect on the halo masses, and thus, on our results regarding 
the halo mass function.

\begin{figure} 
\includegraphics[width=8.7cm]{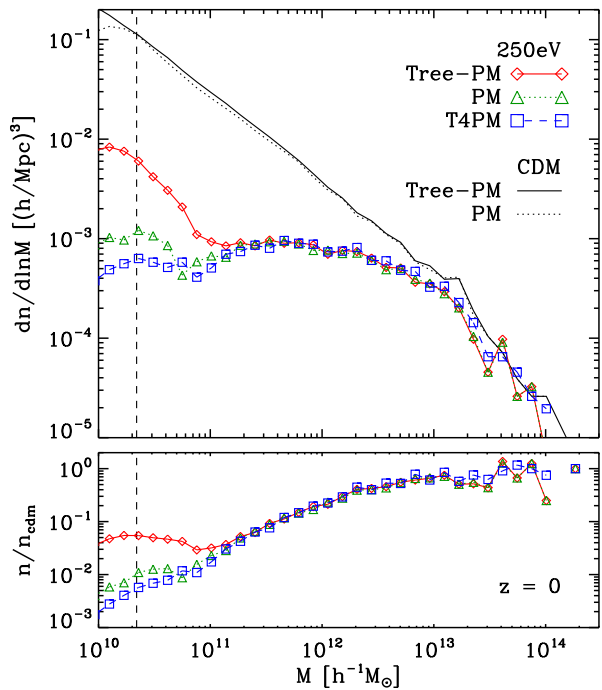} 
\caption{ Comparison of the $z=0$ SO halo mass function in a WDM scenario for a
250eV dark matter particle mass. Coloured lines show the predictions of three
different methods to compute gravitational forces: standard Tree-PM (red line),
only PM (blue line), and the method of Hahn~et~al. (2012) (blue line) We also
display the mass function expected for a CDM case for comparison (black line).
Vertical dashed lines indicate a limit where the abundance of haloes is not
affected by finite force resolution. The bottom panel shows the ratio of our
results to the expectations for a CDM particle.  \label{fig:mf}} 
\end{figure}

\begin{figure*} 
\hspace*{-2.cm}\includegraphics[width=20cm]{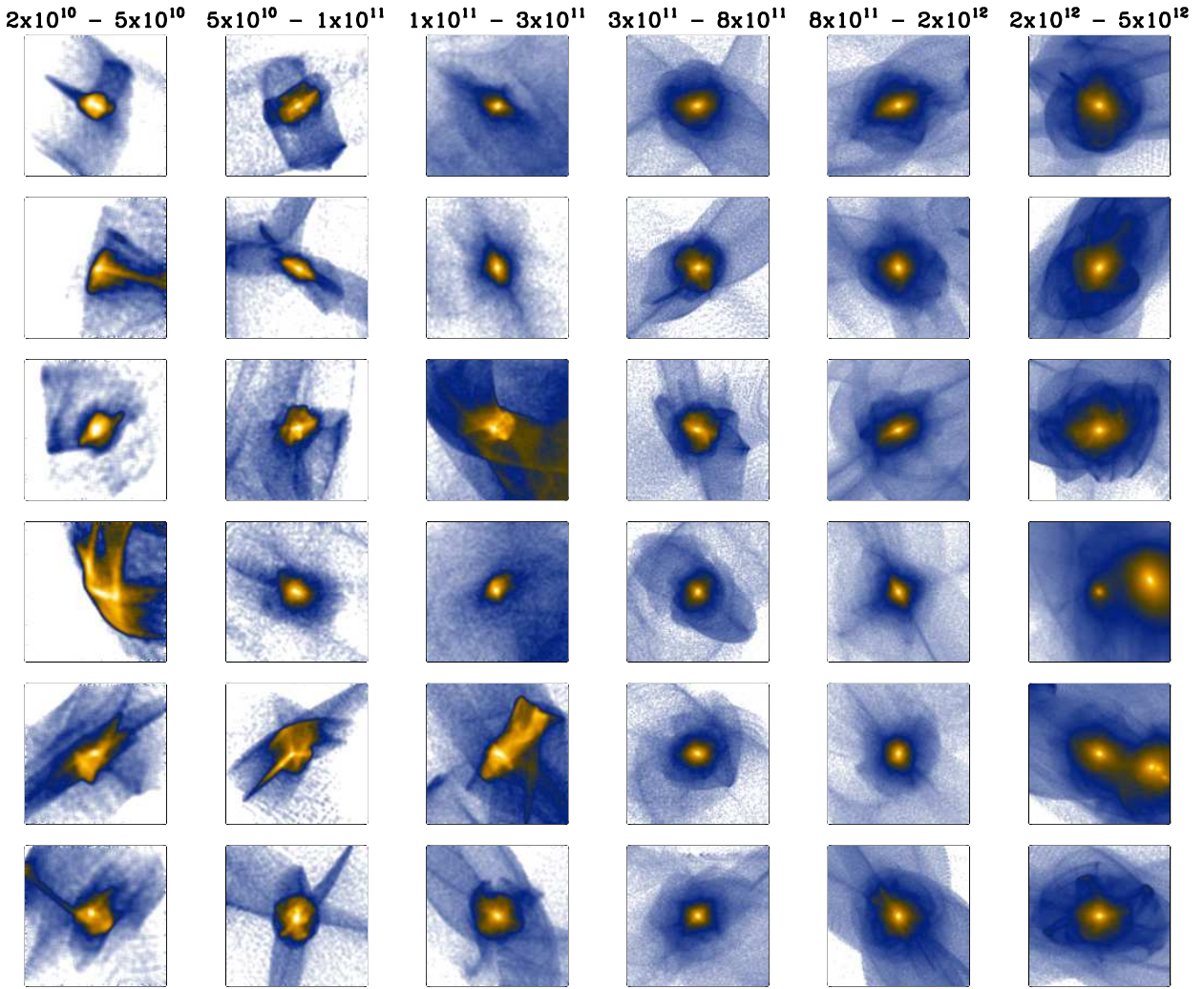}  
\vspace*{-1.5cm}
\caption{ Images of randomly-chosen objects in six disjoint mass bins.  These mass
bins are equally spaced in log M over the mass range $2\times10^{10} <M_{200}/(\Mass) < 5\times10^{12}$.  
Each image displays the logarithmic
projected density field computed using the {\treat} method.  The extent of each image is
equal to $2\times R_{200}$, i.e. twice the virial radius of the respective halo.
\label{fig:examples}}
\end{figure*}

In Fig.~\ref{fig:mf}, we can quantitatively see the differences in the predicted
number of DM haloes at $z=0$, as a function of their SO mass, $M_{200}$, for our
three methods. For
comparison, we also display the halo mass function analogously constructed for
a CDM simulation with matching volume and mass resolution, and where forces are
computed using the {\tree} and {\pmesh} method. For the latter, we use a mesh
of $2048^3$ cells, matching the force resolution of the {\treat} run. The 
vertical dashed line indicates a mass of $2\times10^{10}\Mass$, or, 
equivalently, $\sim700$ particles. This is an estimate for the mass limit above 
which we expect our results to be numerically robust. 

We choose this limit by comparing the resulting mass function in the CDM 
case for the {\pmesh} and {\tree} force methods. Below $M_{\rm min} = 2\times10^{10}\Mass$,
the {\pmesh} mass function shows a strong deficit of haloes, this is (1) because 
the force resolution of the {\pmesh} run ($80\,\kpc$) is simply too low to resolve (and keep bound) 
some low-mass density peaks, which in the {\tree} run collapse as haloes; and (2) 
since it is plausible that even in CDM the lowest masses are a mixture of artifical
fragments and true haloes, as suggested by the strong increase of the number
of  ``peakless'' haloes at low particle counts \citep[cf.][]{Ludlow2011}. Above 
$M_{\rm min}$ this does not seem to be important and the mass functions are consistent
with each other, showing only a small offset caused by a systematic underestimation 
of masses in the {\pmesh} case. This is an indication that our results
above $M_{\min}$ are numerically robust, and that the differences regarding
artificial fragmentation are a result of our improved estimation of the force field
and the respective reduction of discreteness effects, rather than due to a product
of a somewhat low force resolution. Another aspect supporting this is the
fact that the amount of low-mass haloes in the {\treat} run is {\it lower} than in 
the {\pmesh} case, despite the former having {\it higher} force resolution. This
is the opposite to what is expected if the suppression were caused by a lack of
force resolution.

As can be seen in the lower panel of Fig.~\ref{fig:mf}, above 
$M \sim 3\times10^{11}\,\Mass$ the halo mass function of our WDM three runs 
agree well with each other. The suppression with respect to CDM 
reaches a factor or two at $M = 4\times10^{12}\Mass$, but it
can be as large as a factor of $20$. These values are consistent with previous studies, 
though slightly stronger than those reported by \cite{Schneider2012}, however, 
the discrepancy is most likely due to the different halo definitions. The 
qualitative agreement with other works also is not surprising, given that 
the {\tree} method has been the choice of those studies and it agrees with 
our other two methods. It is when we 
consider smaller masses, were this is 
no longer true, that we can enter into a regime hidden to previous simulations. 

Below $M \sim 3\times10^{11}\,\Mass$, and for over an order of magnitude in halo
mass, the {\treat} and {\pmesh} methods deviate 
systematically from the {\tree} run. The characteristic upturn in the halo mass 
function produced by artificial fragmentation is not present in either the {\pmesh} or 
the {\treat} run. The differences reach a maximum factor of $\sim 10$ and $7$, 
respectively at our mass resolution limit $M\sim2\times10^{10}\Mass$. All of this is
consistent with the qualitative picture provided in Fig.~\ref{fig:field}

On the other hand, despite the lack of spurious fragmentation, there is no 
sign of a sharp cut-off, even in the {\treat} run, as expectations raised 
from previous works suggest \cite[e.g.][]{Benson:2013,Schneider2013}. 
The abundance of low-mass objects only decreases slowly and
shows  a mild upturn at $\sim5\times10^{10}\Mass$. We explore 
this issue in more detail next.

\subsection{The nature of collapsed structures}

In order to explore the nature of the objects below the mass cut-off, and the
origin of the mild upturn at $\sim5\times10^{10}\Mass$, we have
visually inspected {\em all} haloes found above our resolution limit in the {\treat}
run. In Fig.~\ref{fig:examples} we provide density projections of six randomly
chosen objects found in six disjoint mass bins, which serve as examples of
the type of objects that populate different regions of the halo mass function.
We display mass carriers in a region of twice the size of the
virial radius around each  target halo. 
 
It is readily apparent that truly different structures are identified at the different 
mass scales. We now enumerate the most common features found in different mass bins:

\begin{itemize}

\item[1)] In the leftmost column (smallest mass bin), we find density peaks just 
undergoing collapse, with highly disturbed morphologies, irregular boundaries 
and that usually show no clear center. In addition, we also find locally overdense regions
that are however incompatible with the concept of a virialized dark matter halo. Most of these
correspond to caustics -- usually located at the radius where the
particle orbits first turn back after crossing a potential minimum.
Another, somewhat less common, occurrence of these are the centres of
very dense filaments, and also the caustics of filaments.

\item[2)] In the next mass bin, we typically find objects where there is 
clearly one of the three axes that has collapsed recently. These objects usually show a 
roughly round external iso-density contour, and a bar-like feature at their centre,
which is the remnant of the filament whose folding produced the collapse
of the objects. 

\item[3)] Objects in the third mass bin show less strong disturbances. They correspond to
roughly spherically symmetric objects, but they clearly show many caustics, resulting
from the continuous folding of the phase-space sheet. Commonly, they also
show a bar, as those in the previous mass bin. 

\item[4-5-6)] Finally, in the three most massive bins, we find systems similar to those 
we usually find in CDM simulations and that can be unequivocally categorised 
as fully collapsed DM structures, with a well defined centre and approximate
spherical symmetry. The objects have much more clearly undergone an isotropic 
virialization than objects on lower mass scales. 

\end{itemize}

\begin{figure} 
\includegraphics[width=8.5cm]{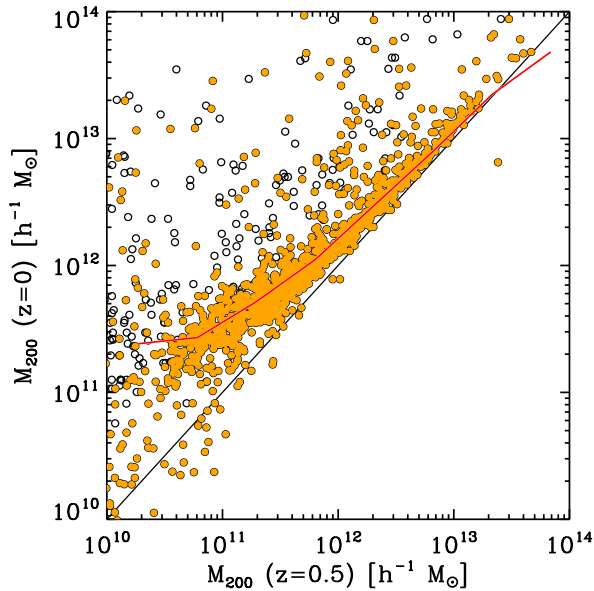} 
\caption{ The relation between halo mass at $z=0.5$ and its descendant 
at $z=0$. The red solid line indicate the median descendant halo mass 
at $z=0$ for ten bins of width $\Delta \log_{10}{M} = 0.5$ in progenitor 
mass at $z=0.5$. Open circles show results for all the haloes at $z=0.5$, 
whereas orange symbols highlight only the most massive progenitor of $z=0$ 
halos. The diagonal black line denotes a 1:1 relation.
\label{fig:desc}} 
\end{figure}

It is interesting to note that the evolutionary state of structures of increasing 
mass resembles different stages of halo formation in Hot dark matter 
cosmologies. Firstly, the local tidal field halts the
expansion of a density perturbation along one axis, which eventually 
collapses, creating a `pancake'. Then, collapse along a second 
orthogonal axes produces a filament. Material is accreted 
along that filament and the third and final axis collapses. Then,
a relaxation process occurs, which finally gives rise to a DM halo 
in the sense of an approximately spherical and virialized object.
Concurrently, the winding of the phase-space sheet  continuously creates
new caustics. Initially, these caustics are strong features, but as new ones
are created and relaxation processes take place, they blend to 
produce a smooth density profile.
If the connection between formation stage and halo mass is true, 
our picture suggests that haloes below the cut-off scale are simply 
haloes during early stages of their formation. This connection is 
also rather natural, as haloes at the cut-off scale should be
forming monolithically as in HDM, not from accreting smaller objects 
as in CDM.
 
This picture is supported by Fig.~\ref{fig:desc} where we show
the descendant halo mass at $z=0$ of haloes detected at $z=0.5$. We link
haloes at these two redshifts by finding the object at $z=0$ that
contains the majority of the particles associated with a FoF halo at
$z=0.5$. We highlight as filled orange symbols the $z=0.5$ haloes that 
are the most massive progenitor of a $z=0$ halo. 

From this figure we can see that almost all haloes increase their
mass consistent with a hierarchical picture of structure
formation.\footnote{We find that the small number of haloes that see a reduction
of their mass are systems that were accreted by a larger halo, their
outskirts removed by tidal stripping, but the denser core survives in a orbit
that yield them to outside the virial radius of the host halo, and thus are
identified as separate haloes but with a reduced mass.}  However,
haloes of different initial masses grow by significantly different amounts. 
Those of $10^{13}\,\Mass$ increase their mass by 30\%, on average. At the
expected half-mode mass scale, $2\times10^{12}\Mass$, the increment
is typically a factor of $2$. Whereas, at $10^{11}\,\Mass$ it is a factor
of 15! This very rapid mass increase at low masses has the consequence that most of the haloes 
below $10^{11}\Mass$ (e.g. those in the first two columns of 
Fig.~\ref{fig:examples}) are well above this limit by $z=0$. This is consistent 
with the picture given above in which the objects found below the mass scale 
are simply a transient stage of halo formation, thus they quickly increase 
their mass and sit above the cut-off scale, where the growth proceeds at a 
slower pace. This is to be contrasted with the CDM case, where low-mass
haloes have the earliest formation redshifts.

Out of the $1413$ points we display, there are $47$ which are located 
below $1\times10^{11}\Mass$ at both $z=0.5$ and $z=0.0$. These haloes are
not compatible with our interpretation, and they could be rare occasion
where our mass resolution is not sufficient to avoid absolutely all
fragmentation. Despite this, this is a very small population, which will
not affect our results.

\subsection{The abundance of virialized structures}

With the ideas discussed above in hand, we now return to the issue of the halo
mass function in WDM cosmologies. Upon visual inspection of the members of
our FoF-SO catalogue, it became obvious that many of the entries did 
not comply with the features usually found in halo catalogues built
from CDM simulations. Consequently, we visually inspected and classified 
all haloes in our {\treat} run into one of three groups. 

\begin{itemize}
\item[1.] {\bf ``Not Halos''}: In the first category we include all objects 
that appear as clear failures of our halo finder algorithm. These enclose 
mainly three cases: one corresponds to outer caustics of large haloes (which
sometimes are located further than the virial radius), another to 
descendants of haloes that have flown trough a more massive system (these are
stripped of most of their mass, but their core survives). The third case
correspond to dense sheets and filaments where sometimes the collapse of a 
further axis has started. 

\item[2.] {\bf ``Proto-Halos''} Our second category contains haloes that are 
not fully formed yet, but show clear isolated 3D density enhancements. Here 
all three axes have collapsed, but the density peak has not fully virialised: 
we include here all objects from highly 
anisotropic systems, that appear just after a violent collapse, to much more
quiet haloes, where only minor departures from a smooth mass distribution exist.

\item[3.] {\bf ``Halos''} The third and final category contains systems which 
can be unambiguously defined as a halo in the traditional sense of approximately 
spherical objects showing clear three dimensional virialization, and that resemble
those seen in CDM simulations.
\end{itemize}

We note that we attempted to perform automatic classifications using several
different halo properties. Unfortunately, none of them could satisfactory separate 
the categories mentioned above. Some of the measures introduced in \cite{Abel2012}
or methods inspecting the shape of the Lagrangian patch of the identified structures 
\citep[such as that proposed by e.g.][]{Lovell2012} may help in automating 
such a procedure. On the other hand, since proto-haloes, as well as some ``Not Halo'' 
objects, are likely simply early stages of halo formation, they are also likely to 
correspond to  peaks in the initial conditions with only their collapse time
differing from those corresponding to ``Halos''. Thus, Lagrangian approaches might 
not clearly separate our three classes of objects. An additional complication for
automatic classifications is that the haloes in the critical regime, $<10^{11}\Mass$,
are resolved with only a few thousand particles which is not enough to perform a
detailed analysis of their internal structure. We will defer further exploration 
of these issues to future work.

\begin{figure} 
\includegraphics[width=8.7cm]{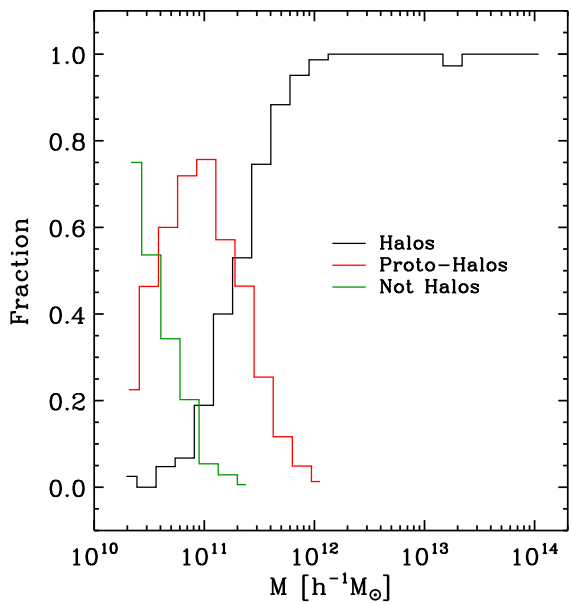} 
\caption{ Relative contribution of different types of
objects to the total WDM halo mass function. The black
histogram shows the fraction of ``Halos''. The green
histogram shows the abundance of ``'Proto-Haloes', whereas
the green line indicates the fraction of objects that our
SO-FOF algorithm wrongly identified as haloes. See the text
for more details about our classification method.
\label{fig:fraction}}
\end{figure} 

\begin{figure} 
\includegraphics[width=8.7cm]{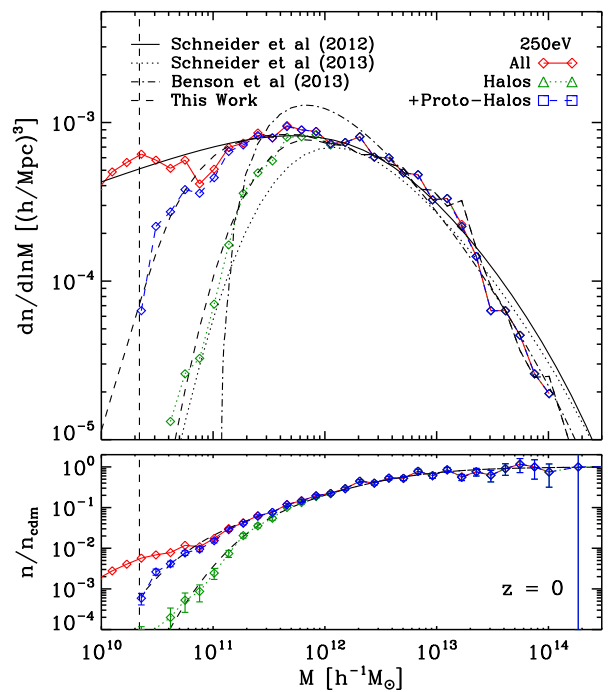} 
\caption{ Contribution of different types of
objects to the WDM halo mass function. The red line show
the abundance of standard dark matter haloes. The green line represents haloes
in final stages of formation, while the blue line displays the abundance of
objects in initial stages of formation. Finally, the magenta line shows objects
incorrectly identified as haloes by our algorithm. See the text for more
details on the classification, and Fig.~\ref{fig:examples} for examples of
structures in the various categories.  Black lines of different styles show the
mass function expected in different analytical formulations, as specified in
the legend.  Vertical dashed line indicate a limit where the abundance of
haloes is not affected by finite force resolution. The bottom panel shows the
ratio of our results to the expectations in a CDM scenario.
\label{fig:mf_wdm}} 
\end{figure}

In Fig.~\ref{fig:fraction}, we show the relative contribution of each of these
three categories to the WDM halo mass function.  It is very interesting to see
that the groups are clearly localized at different mass scales, although some
overlap exists. High masses are dominated by standard haloes.  Right below the
cut-off, recently collapsed systems dominate. And the lowest masses receive a
similar contribution from {\proto} and from failures of our FoF-SO algorithm. 

Before we continue, we would like to note that, as in most classifications, the
division between these three groups is somewhat arbitrary. This is accentuated
by the subjective nature of our visual inspection.  For these reasons, we
emphasise that the distinction between different categories just provides a
qualitative assessment of the nature of objects at different mass scales, and
of how they affect the WDM halo mass function. 

Another point to note is that the fine division between the categories does
depend on the force resolution employed. We have explored this by carrying out
our {\treat} run using a PM mesh a factor of eight smaller, thus degrading our
spatial resolution by a factor of two. There are three aspects worth noting. 

1) The amount of filaments/sheets in our catalogues, as well as the sum of
{\halo} plus {\proto} remains roughly the same when the force resolution is
varied. This is because the time of collapse of a filament depends mostly on
large-scale tidal and density fields, which are less sensitive to the force
resolution. 

2)  The distinction  between {\halo} and {\proto} is very different at
different force resolutions. With higher force resolution, caustics are created
more rapidly, there is more mixing, and haloes appear more relaxed. Note that
due to computational limitations, it is not possible for us to increase the
mass resolution of our runs needed to increase further the force resolution.

3) The frequency of some type of FoF failures changes considerably with force
resolution.  In this case, we find that the number of {\nohalo} at the low mass
end increased substantially, mainly due to an increase in the number of
caustics -- a lower force resolution allows the turn-around radius to move
outwards. 

{\em From this, we can conclude that the mass functions for the {\halo} and the
{\halo} plus {\proto} samples should provide the range in which we expect the
mass function of virialised haloes to lie.} 

We are now in the position to provide the most important result of our paper.
In Fig.~\ref{fig:mf_wdm}, we show the WDM halo mass function for two
catalogues.  The first one, denoted by a green dotted line, shows the abundance
of systems categorised as {\halo}, the second case, denoted by a blue dashed
line, adds in the contribution of {\proto}. Thus, the blue and green lines
provide upper and lower limits to the abundance of collapsed objects in WDM
cosmologies.  These data are well fitted by the following functional form:

\begin{equation}
\frac{n}{n_{\rm cdm}}(M) = \frac{1}{2} \left(1 + \frac{M_1}{M} \right)^{-1}  
               \left[ 1 + {\rm erf}\left(\log{\frac{M}{M_2}}\right) \right]
\end{equation}

\noindent with $M_1 = 3.9\times10^{12}\Mass$ set by the half-mode mass scale,
and $M_2$ corresponds to the location of the small-scale cut-off, which we find
to be  $5.2\times10^{10}$ for the {\halo} catalogue, and and
$2.1\times10^{11}\Mass$ for the {\halo} plus {\proto} sample. The best fits are
displayed as dashed lines in Fig.~\ref{fig:mf_wdm}. 

We clearly see that once we neglect the contribution of failures of our halo
finder, the WDM halo mass function shows a strong cut-off at small masses, and
the upturn seen in Fig.~\ref{fig:mf} essentially disappears. The strong cut-off
implies that there are no collapsed peaks below some scale, in agreement with
what one would naively expect from the cut-off in the transfer function which
predicts no small-scale density perturbations. This rules out the scenario in
which a substantial population of low-mass haloes are created as a result of
nonlinear self-interactions of the density field.  However, a quantitative
comparison between the cut-off in the transfer function and that in the mass
function is not straightforward, as the substantial differences between our two
samples indicate. 

To illustrate this point, in Fig.~\ref{fig:mf_wdm}, we also show three
predictions for the shape of the WDM halo mass function. Dotted line shows the
recent results of \cite{Schneider2013} that is based on a sharp-k filter
calibrated with a {\tree} simulation, whereas the solid line is the prediction
of \cite{Schneider2012} based on the EPS formalism and extrapolating the
results of N-body simulations. Dash-dotted line is the model of
\cite{Benson:2013}, who attemp to incorporate the effects of the thermal
velocity of the WDM particle. All predictions differ largely.  The
\cite{Schneider2012} prediction, for instance,  does not predict any cut-off in
the halo mass function, whereas the other two models do so and roughly match
the {\halo} sample. However, as we discussed below, our data only provides a
lower limit for the halo mass function, and the differences in the halo
definition become important for a quantitative comparison. 

The lack of an unique answer is a natural consequence of the complexity of
structure formation and of the need for a common definition of a ``halo'' when
numerical simulations and analytical formalism are compared. This is also true
in CDM, but accentuated in WDM since low-mass scales are dominated by the
collapse of filaments and by systems in the process of relaxation or collapse.
Despite the added complexity, examining the cut-off scale can be extremely
useful to isolate successful as well as problematic aspects of analytic
formulations, which in turn could lead to improvements in their foundations
and ultimately to a better understanding of structure formation and assembly in
all cosmological models of structure formation.

\section{Conclusions}

In this paper, we were able to overcome the notorious difficulties associated
with numerical simulations of the evolution of warm dark matter models. For
decades, the numerical fragmentation of filaments created an artificial
population of low mass haloes which dominated the halo mass function on small
mass scales. Here, we demonstrated that it is possible to avoid such artefacts
by employing a force resolution consistent with the mass resolution of the
simulations. With this, and for the first time, we could explore the halo mass
function below the cut-off scale. We discovered a picture more complex than
that present in CDM simulations. 

Structure formation in scenarios that have a small scale cut off in the power
spectrum proceeds quite differently than in the CDM model. In CDM, haloes form
mainly from accreting smaller haloes, which have the earliest formation times
and the longest time to relax. On the contrary, in WDM, a large fraction of
haloes forms from the direct collapse of filaments and small-mass haloes are
typically the most recently formed, and thus are still in a process of
virialisation.

We find that this formation mechanism is imprinted in the WDM halo mass
function: essentially different stages of halo formation dominate the counts at
different masses. On the smallest mass scales we can resolve, identified
objects are typically centres of  filaments that are starting to collapse. On
intermediate mass scales, objects typically correspond to fluctuations that
have collapsed and are in the process of relaxation, whereas the high mass end
is dominated by objects similar to haloes identified in CDM simulations.

In addition, we found that traditional group-finders produce, on small mass
scales, catalogues with objects not consistent with the definition of a halo.
In a CDM calculation, essentially all dense structures are always part of
haloes, and even the simplest approaches, such as the FoF algorithm, succeed in
selecting appropriate objects above a certain mass limit
\citep[e.g.][]{Warren2006, More2011}. On the other hand, in WDM these
approaches prove to be unreliable. The small-scale cut-off in the primordial
transfer function, implies that there are dense structures (e.g. filaments,
caustics of large haloes) that can be incorrectly considered as haloes. These
misidentifications dominate the WDM mass function below the cut-off scale.  For
the simulations we considered here, we could bypass this problem by visually
inspecting and classifying  different systems. This is not prohibitively
demanding, thanks to the relatively low number of systems in our runs.
However, of course, this is in general not true, and improved halo finders are
desirable for future work. 

After neglecting these failures of halo identification, we observe a strong
cut-off in the halo mass function, with very few objects found below the
cut-off scale. These correspond to haloes undergoing rapid collapse and
virialisation. Our results indicate that the cutoff scale in the initial
density fluctuations does indeed translate into a comparably strong cutoff
scale in the halo mass function~(Fig.~\ref{fig:mf_wdm}).  This implies that
filaments and sheets that formed in early stages of gravitational collapse are
remarkably stable and do not fragment due to the lack of small scale
perturbations. 

Our work poses several questions. The most natural one is the role of
artificial fragmentation in the internal properties of DM haloes. In the
absence of spurious haloes, filaments get denser and the accretion of material
happens continuously from filaments, not as a sequential accretion of small
dense haloes which would likely behave differently dynamically. It is not clear
whether this difference will impact, for instance, the concentration of haloes
near the cut-off scale. Thus, we plan to carry out a suite of higher-resolution
simulations to address this.

Another question arises from the fact that the exact mass scale at which the
cut-off in the halo mass function is located, does depend on the exact
definition of what constitutes a halo. Especially, it depends on the separation
between a fully formed halo and a halo in the process of formation and/or
relaxation.  A priori it is thus not clear what one may wish to call a halo.
Functional definitions might involve regions of sufficiently high DM density
that would allow the associated baryons to collapse and cool, or surface
densities that allow for gravitational lensing or more theoretically inspired
measures such as a limit on the velocity anisotropy of the dark matter
particles or perhaps particular axis ratios inferred from the velocity
ellipsoid or inertial tensor. Whatever the exact definition one may choose the
abundance of ``haloes'' fulfilling these choices will likely vary quite
dramatically.

Therefore, any key ingredient for a new halo finder will be related directly to
the  properties that one aims to select for. For instance, the isotropy of the
velocity dispersion, or the number of streams, and include the requirement that
all three axes have collapsed \citep{Falck2012} perhaps augmented with local
measures of the velocity dispersion in all three directions~\citep{Abel2012}.
A common definition is also needed for a proper comparison with analytic models
for structure formation. It will be desirable for future work to investigate
whether there are generic definitions that sensibly define a concept of dark
matter halo that is applicable to CDM as well as WDM computational cosmology
questions. 

Even once the issue of halo definition is decided upon, there is another
problem related to the numerical methods. We find that the exact virialisation
state as well as the number of caustics depend sensitively on the force
resolution employed in our simulations. This is different from standard CDM
runs because of the difference in the formation mechanism, and because of a
special combination of force and mass resolution. However, this also opens an
exciting possibility: because of the free-streaming scale in WDM, there is also
a upper limit to the density and also a scale for small-scale gravitational
interactions, which could eventually allow us to simulate the full range of
length and mass scales relevant for the formation of a halo.

The question of halo definition, and the role of artificial fragmentation, is
closely related to the issue of what is the minimum halo mass and/or
virialisation stage for a halo to host galaxies. Fragmentation of gas into
stars may well still occur in places not regarded as haloes in the traditional
sense, perhaps already in regions of filamentary collapse: anywhere where a
local gravitational potential well is already aggregating the baryons. On the
contrary, the continuous folding of the phase-space sheet might continuously
shock-heat the gas, and the cut-off in the respective galaxy stellar-mass
function might not trivially relate to that in the halo mass function. 

Unfortunately, given the strong limits on the mass of the potential WDM
particle from Lyman-$\alpha$ forest constraints \cite{Viel2005,Boyarsky2009}
the mass scale on which such a different galaxy formation scenario could be
related to the observable Universe is severely limited, and other physics might
be more relevant.  Nevertheless, WDM scenarios remain of great interest to
obtain a much better theoretical understanding of how dark matter haloes
assemble and how the collisionless fluid virializes. In particular, they very
closely resemble the events that lead to the very first dark matter haloes even
in a CDM scenario albeit on radically different mass and length scales
\citep{Diemand2005b, Goerdt2007,Ishiyama2010}. We foresee our results
stimulating a more detailed exploration of the formation of the smallest
structures expected to form in a given cosmology, which will, hopefully,
advance our overall understanding of structure formation.

\section*{Acknowledgements} 
We acknowledge useful discussions with Yu Lu, Aaron Ludlow, Ralf Kaehler, Aseem
Paranjape, and Jesus Zavala. We thank A. Benson for providing his model data 
for our Fig.~7. OH acknowledges support from the Swiss National Science
Foundation (SNSF) through the Ambizione fellowship. TA acknowledges support by
LDRD program at the SLAC National Accelerator Laboratory as well as the Terman
fellowship at Stanford University.  We gratefully acknowledge the support of
Stuart Marshall and the SLAC computational team, as well as the computational
resources at SLAC.

\bibliographystyle{mn2e} \bibliography{database}

\begin{thebibliography}{50}
\expandafter\ifx\csname natexlab\endcsname\relax\def\natexlab#1{#1}\fi

\bibitem[{{Abel} {et~al}\mbox{.}(2012){Abel}, {Hahn}, \& {Kaehler}}]{Abel2012}
{Abel} T., {Hahn} O., {Kaehler} R., 2012, \mnras, 427, 61

\bibitem[{{Angulo} {et~al}\mbox{.}(2013){Angulo}, {Hahn}, \&
  {Abel}}]{Angulo2013}
{Angulo} R.~E., {Hahn} O., {Abel} T., 2013, ArXiv e-prints

\bibitem[{{Angulo} {et~al}\mbox{.}(2012){Angulo}, {Springel}, {White},
  {Jenkins}, {Baugh}, \& {Frenk}}]{Angulo2012a}
{Angulo} R.~E., {Springel} V., {White} S.~D.~M., {Jenkins} A., {Baugh} C.~M.,
  {Frenk} C.~S., 2012, \mnras, 426, 2046

\bibitem[{{Avila-Reese} {et~al}\mbox{.}(2001){Avila-Reese}, {Col{\'{\i}}n},
  {Valenzuela}, {D'Onghia}, \& {Firmani}}]{Avila-Reese:2001}
{Avila-Reese} V., {Col{\'{\i}}n} P., {Valenzuela} O., {D'Onghia} E., {Firmani}
  C., 2001, \apj, 559, 516

\bibitem[{{Barnes} \& {Hut}(1986)}]{Barnes1986}
{Barnes} J., {Hut} P., 1986, \nat, 324, 446

\bibitem[{{Benson} {et~al}\mbox{.}(2013){Benson}, {Farahi}, {Cole},
  {Moustakas}, {Jenkins}, {Lovell}, {Kennedy}, {Helly}, \&
  {Frenk}}]{Benson:2013}
{Benson} A.~J. {et~al.}, 2013, \mnras, 428, 1774

\bibitem[{{Bode} {et~al}\mbox{.}(2001){Bode}, {Ostriker}, \&
  {Turok}}]{Bode2001}
{Bode} P., {Ostriker} J.~P., {Turok} N., 2001, \apj, 556, 93

\bibitem[{{Bond} \& {Szalay}(1983)}]{Bond1983}
{Bond} J.~R., {Szalay} A.~S., 1983, \apj, 274, 443

\bibitem[{{Boyarsky} {et~al}\mbox{.}(2009{\natexlab{a}}){Boyarsky},
  {Lesgourgues}, {Ruchayskiy}, \& {Viel}}]{Boyarsky2009}
{Boyarsky} A., {Lesgourgues} J., {Ruchayskiy} O., {Viel} M.,
  2009{\natexlab{a}}, \jcap, 5, 12

\bibitem[{{Boyarsky} {et~al}\mbox{.}(2009{\natexlab{b}}){Boyarsky},
  {Ruchayskiy}, \& {Shaposhnikov}}]{Boyarsky2009a}
{Boyarsky} A., {Ruchayskiy} O., {Shaposhnikov} M., 2009{\natexlab{b}}, Annual
  Review of Nuclear and Particle Science, 59, 191

\bibitem[{{Boylan-Kolchin} {et~al}\mbox{.}(2011){Boylan-Kolchin}, {Bullock}, \&
  {Kaplinghat}}]{Boylan-Kolchin2011}
{Boylan-Kolchin} M., {Bullock} J.~S., {Kaplinghat} M., 2011, \mnras, 415, L40

\bibitem[{{Centrella} {et~al}\mbox{.}(1988){Centrella}, {Gallagher}, {Melott},
  \& {Bushouse}}]{Centrella1988}
{Centrella} J.~M., {Gallagher}, III J.~S., {Melott} A.~L., {Bushouse} H.~A.,
  1988, \apj, 333, 24

\bibitem[{{Col{\'{\i}}n} {et~al}\mbox{.}(2000){Col{\'{\i}}n}, {Avila-Reese}, \&
  {Valenzuela}}]{Colin2000}
{Col{\'{\i}}n} P., {Avila-Reese} V., {Valenzuela} O., 2000, \apj, 542, 622

\bibitem[{{Col{\'{\i}}n} {et~al}\mbox{.}(2008){Col{\'{\i}}n}, {Valenzuela}, \&
  {Avila-Reese}}]{Colin2008}
{Col{\'{\i}}n} P., {Valenzuela} O., {Avila-Reese} V., 2008, \apj, 673, 203

\bibitem[{{Colombi} {et~al}\mbox{.}(1996){Colombi}, {Dodelson}, \&
  {Widrow}}]{Colombi1996}
{Colombi} S., {Dodelson} S., {Widrow} L.~M., 1996, \apj, 458, 1

\bibitem[{{Dalcanton} \& {Hogan}(2001)}]{Dalcanton2001}
{Dalcanton} J.~J., {Hogan} C.~J., 2001, \apj, 561, 35

\bibitem[{{Davis} {et~al}\mbox{.}(1985){Davis}, {Efstathiou}, {Frenk}, \&
  {White}}]{Davis1985}
{Davis} M., {Efstathiou} G., {Frenk} C.~S., {White} S.~D.~M., 1985, \apj, 292,
  371

\bibitem[{{Diemand} {et~al}\mbox{.}(2005){Diemand}, {Moore}, \&
  {Stadel}}]{Diemand2005b}
{Diemand} J., {Moore} B., {Stadel} J., 2005, \nat, 433, 389

\bibitem[{{Efstathiou} \& {Eastwood}(1981)}]{Efstathiou1981}
{Efstathiou} G., {Eastwood} J.~W., 1981, \mnras, 194, 503

\bibitem[{{Eisenstein} \& {Hu}(1999)}]{EisensteinHu1999}
{Eisenstein} D.~J., {Hu} W., 1999, \apj, 511, 5

\bibitem[{{Falck} {et~al}\mbox{.}(2012){Falck}, {Neyrinck}, \&
  {Szalay}}]{Falck2012}
{Falck} B.~L., {Neyrinck} M.~C., {Szalay} A.~S., 2012, \apj, 754, 126

\bibitem[{{Goerdt} {et~al}\mbox{.}(2007){Goerdt}, {Gnedin}, {Moore}, {Diemand},
  \& {Stadel}}]{Goerdt2007}
{Goerdt} T., {Gnedin} O.~Y., {Moore} B., {Diemand} J., {Stadel} J., 2007,
  \mnras, 375, 191

\bibitem[{{Hahn} \& {Abel}(2011)}]{HahnAbel2011}
{Hahn} O., {Abel} T., 2011, \mnras, 415, 2101

\bibitem[{{Hahn} {et~al}\mbox{.}(2012){Hahn}, {Abel}, \& {Kaehler}}]{Hahn2012}
{Hahn} O., {Abel} T., {Kaehler} R., 2012, ArXiv e-prints

\bibitem[{{Hockney} \& {Eastwood}(1981)}]{Hockney1981}
{Hockney} R.~W., {Eastwood} J.~W., 1981, {Computer Simulation Using Particles}.
  Computer Simulation Using Particles, New York: McGraw-Hill, 1981

\bibitem[{{Ishiyama} {et~al}\mbox{.}(2010){Ishiyama}, {Makino}, \&
  {Ebisuzaki}}]{Ishiyama2010}
{Ishiyama} T., {Makino} J., {Ebisuzaki} T., 2010, \apjl, 723, L195

\bibitem[{{Kaehler} {et~al}\mbox{.}(2012){Kaehler}, {Hahn}, \&
  {Abel}}]{Kaehler2012}
{Kaehler} R., {Hahn} O., {Abel} T., 2012, ArXiv e-prints

\bibitem[{{Knebe} {et~al}\mbox{.}(2003){Knebe}, {Devriendt}, {Gibson}, \&
  {Silk}}]{Knebe2003}
{Knebe} A., {Devriendt} J.~E.~G., {Gibson} B.~K., {Silk} J., 2003, \mnras, 345,
  1285

\bibitem[{{Komatsu} {et~al}\mbox{.}(2010){Komatsu}, {Smith}, {Dunkley},
  {Bennett}, {Gold}, {Hinshaw}, {Jarosik}, {Larson}, {Nolta}, {Page},
  {Spergel}, {Halpern}, {Hill}, {Kogut}, {Limon}, {Meyer}, {Odegard}, {Tucker},
  {Weiland}, {Wollack}, \& {Wright}}]{Komatsu2010}
{Komatsu} E. {et~al.}, 2010, ArXiv e-prints

\bibitem[{{Lacey} \& {Cole}(1993)}]{Lacey1993}
{Lacey} C., {Cole} S., 1993, \mnras, 262, 627

\bibitem[{{Lovell} {et~al}\mbox{.}(2012){Lovell}, {Eke}, {Frenk}, {Gao},
  {Jenkins}, {Theuns}, {Wang}, {White}, {Boyarsky}, \&
  {Ruchayskiy}}]{Lovell2012}
{Lovell} M.~R. {et~al.}, 2012, \mnras, 420, 2318

\bibitem[{{Ludlow} \& {Porciani}(2011)}]{Ludlow2011}
{Ludlow} A.~D., {Porciani} C., 2011, \mnras, 413, 1961

\bibitem[{{Macci{\`o}} {et~al}\mbox{.}(2013){Macci{\`o}}, {Ruchayskiy},
  {Boyarsky}, \& {Mu{\~n}oz-Cuartas}}]{Maccio2013}
{Macci{\`o}} A.~V., {Ruchayskiy} O., {Boyarsky} A., {Mu{\~n}oz-Cuartas} J.~C.,
  2013, \mnras, 428, 882

\bibitem[{{Melott}(2007)}]{Melott2007}
{Melott} A.~L., 2007, ArXiv e-prints

\bibitem[{{Melott} \& {Shandarin}(1989)}]{Melott1989}
{Melott} A.~L., {Shandarin} S.~F., 1989, \apj, 343, 26

\bibitem[{{Menci} {et~al}\mbox{.}(2012){Menci}, {Fiore}, \&
  {Lamastra}}]{Menci2012}
{Menci} N., {Fiore} F., {Lamastra} A., 2012, \mnras, 421, 2384

\bibitem[{{More} {et~al}\mbox{.}(2011){More}, {Kravtsov}, {Dalal}, \&
  {Gottl{\"o}ber}}]{More2011}
{More} S., {Kravtsov} A.~V., {Dalal} N., {Gottl{\"o}ber} S., 2011, \apjs, 195,
  4

\bibitem[{{Moroi} {et~al}\mbox{.}(1993){Moroi}, {Murayama}, \&
  {Yamaguchi}}]{Moroi1993}
{Moroi} T., {Murayama} H., {Yamaguchi} M., 1993, Physics Letters B, 303, 289

\bibitem[{{Obreschkow} {et~al}\mbox{.}(2013){Obreschkow}, {Power}, {Bruderer},
  \& {Bonvin}}]{Obreschkow2013}
{Obreschkow} D., {Power} C., {Bruderer} M., {Bonvin} C., 2013, \apj, 762, 115

\bibitem[{{Press} \& {Schechter}(1974)}]{Press1974}
{Press} W.~H., {Schechter} P., 1974, \apj, 187, 425

\bibitem[{{Schneider} {et~al}\mbox{.}(2012){Schneider}, {Smith}, {Macci{\`o}},
  \& {Moore}}]{Schneider2012}
{Schneider} A., {Smith} R.~E., {Macci{\`o}} A.~V., {Moore} B., 2012, \mnras,
  424, 684

\bibitem[{{Schneider} {et~al}\mbox{.}(2013){Schneider}, {Smith}, \&
  {Reed}}]{Schneider2013}
{Schneider} A., {Smith} R.~E., {Reed} D., 2013, ArXiv e-prints

\bibitem[{{Smith} \& {Markovic}(2011)}]{Smith:2011}
{Smith} R.~E., {Markovic} K., 2011, \prd, 84, 063507

\bibitem[{{Splinter} {et~al}\mbox{.}(1998){Splinter}, {Melott}, {Shandarin}, \&
  {Suto}}]{Splinter1998}
{Splinter} R.~J., {Melott} A.~L., {Shandarin} S.~F., {Suto} Y., 1998, \apj,
  497, 38

\bibitem[{{Springel}(2005)}]{Springel2005b}
{Springel} V., 2005, \mnras, 364, 1105

\bibitem[{{Viel} {et~al}\mbox{.}(2005){Viel}, {Lesgourgues}, {Haehnelt},
  {Matarrese}, \& {Riotto}}]{Viel2005}
{Viel} M., {Lesgourgues} J., {Haehnelt} M.~G., {Matarrese} S., {Riotto} A.,
  2005, \prd, 71, 063534

\bibitem[{{Wang} \& {White}(2007)}]{Wang2007}
{Wang} J., {White} S.~D.~M., 2007, \mnras, 380, 93

\bibitem[{{Warren} {et~al}\mbox{.}(2006){Warren}, {Abazajian}, {Holz}, \&
  {Teodoro}}]{Warren2006}
{Warren} M.~S., {Abazajian} K., {Holz} D.~E., {Teodoro} L., 2006, \apj, 646,
  881

\bibitem[{{White} \& {Vogelsberger}(2009)}]{White2009}
{White} S.~D.~M., {Vogelsberger} M., 2009, \mnras, 392, 281

\bibitem[{{Zavala} {et~al}\mbox{.}(2009){Zavala}, {Jing}, {Faltenbacher},
  {Yepes}, {Hoffman}, {Gottl{\"o}ber}, \& {Catinella}}]{Zavala2009}
{Zavala} J., {Jing} Y.~P., {Faltenbacher} A., {Yepes} G., {Hoffman} Y.,
  {Gottl{\"o}ber} S., {Catinella} B., 2009, \apj, 700, 1779

\end{thebibliography}

\label{lastpage} \end{document}